\documentclass{ws-mpla}
\usepackage[super,compress]{cite}
\usepackage{graphicx}
\usepackage{color}
\usepackage{xcolor}

\begin{document}

\markboth{Authors' names}{Instructions for typing manuscripts (paper's title)}

\catchline{}{}{}{}{}

\title{Quantum forgery attacks against OTR structures based on Simon’s algorithm
}

\author{Wenjie Liu\footnote{
Corresponding Author.}}

\address{School of Computer and Software, Nanjing University of Information Science  and Technology, \\
Nanjing, 210044, Jiangsu, China\\
Engineering Research Center of Digital Forensics, Ministry of Education, Nanjing University of Information Science and Technology,\\
Nanjing, 210044, Jiangsu, China\\
wenjiel@163.com}

\author{Mengting Wang}

\address{School of Computer and Software, Nanjing University of Information Science and Technology, \\
Nanjing, 210044, Jiangsu, China\\
}

\author{Zixian Li}

\address{School of Computer and Software, Nanjing University of Information Science  and Technology, \\
Nanjing, 210044, Jiangsu, China\\
}

\maketitle

\pub{Received (Day Month Year)}{Revised (Day Month Year)}

\begin{abstract}
Classical forgery attacks against Offset Two-round (OTR) structures require some harsh conditions, such as some plaintext and ciphertext pairs need to be known, and the success probability is not too high. To solve these problems, a quantum forgery attack on OTR structure using Simon's algorithm is proposed. The attacker intercept the ciphertext-tag pair $(C,T)$ between the sender and receiver, while Simon's algorithm is used to find the period of the tag generation function in OTR, then we can successfully forge new ciphertext $C'$ ($C'\ne C$) for intercepted tag $T$. For a variant of OTR structure (Prøst-OTR-Even-Mansour structure), a universal forgery attack, in which it is easy to generate the correct tag of any given message if the attacker is allowed to change a single block in it, is proposed. It first obtains the secret parameter $L$ using Simon's algorithm, then the secret parameter $L$ is used to find the keys $k_1$ and $k_2$, so that an attacker can forge the changed messages. It only needs several plaintext blocks to help obtain the keys to forge any messages.   Performance analysis shows that the query complexity of our attack is $O(n)$, and its success probability is very close to 1.  

\end{abstract}

\ccode{PACS Nos.: 03.67.Dd.}

\section{Introduction}	

In terms of cryptographic security research, the authentication encryption algorithm can realize the confidentiality and integrity verification of information at the same time, and it has been widely used in various network security systems. The authentication encryption working mode is a cryptographic scheme that encrypts messages to generate ciphertext and calculates authentication labels to solve practical problems such as privacy and authenticity of user information. At present, a large amount of information not only needs to be kept confidential during the transmission process, but also needs to be authenticated after the receiver receives the information to ensure the confidentiality, integrity and authenticity of the information during the transmission process. Therefore, it is very necessary to design and study the authentication encryption algorithm. The goal of the CAESAR competition is to identify reliable, efficient, secure, authenticated cryptographic algorithm combinations with unique properties for different application scenarios. A total of 57 algorithms were collected in the initial stage of this encryption competition.

Offset Two-round (OTR) \cite{Parallelizable2014} is an online, one-pass, authenticated encryption block cipher mode that can be processed in parallel for every two consecutive blocks. The OTR mode is similar in structure to the OCB mode \cite{Rogaway2004}, but the OTR mode only uses the forward function of the block cipher for encryption and decryption algorithms. Its instantiation using an AES block cipher with 128 key bits (called AES-OTR) has become a CAESAR candidate. The Prøst-OTR authenticated encryption algorithm (v1.0/1.1) is also a CAESAR candidate submitted by Kavun {\it et al. }\cite{Prøstv1,Prøstv1.1} inspired by OTR. It incorporates a newly designed efficient permutation, the Prøst permutation. The Prøst-OTR variant (Prøst-OTR-Even-Mansour) uses Prøst arrangement in a single-key Even-Mansour structure as block cipher\cite{Christoph2015}. The Even-Mansour method \cite{Biryukov2000, Bogdanov2012,1991Limitations,Dunkelman2012,Gentry2004,Lampe2012} has been extensively studied, and it has been proven to be secure under different security concepts. Besides, detailed security level and key length bounds are also given, but it is inherently vulnerable to related key. OTR structure is an authentication encryption structure, which can ensure the confidentiality and integrity of information at the same time, and has certain research significance. There are also many scholars who are constantly studying forgery attacks on these two structures. Christoph {\it et al.} \cite{Christoph2015} suggested that the related key properties constructed by Even-Mansour are not well covered by classical security concepts, and they can lead to powerful forgery attacks on Prøst-OTR-Even-Mansour structure. Hassan {\it et al.} \cite{Mahri2016} showed that some primitive polynomials can cause collisions between the masking coefficients used in the current instantiation, allowing forgery on OTR structure. However, as far as we know, these forgery attacks have harsh constraints, which makes the forgery scenario very strict.

On the other hand, in the quantum world, since the Shor algorithm \cite{Shor1999} was proposed, it has been announced that quantum computers will pose a serious threat to public key cryptography. More and more researchers have begun to use quantum algorithms to crack symmetric cryptosystems, such as Simon's algorithm \cite{Kuwakado2010,Kuwakado2012,Simon1997,Kaplan2016,Shi2018}, Grover's algorithm \cite{Grover1997,2017Grover} and Bernstein-Vazirani algorithm \cite{1997Quantum,Xie2019}. In addition, they also proposed some new quantum algorithms \cite{Chailloux2017,Hosoyamada2019}, and even extended classical cryptanalysis methods to the quantum domain \cite{Bonnetain2020, Hosoyamada2018}. It is worth mentioning that there have also been new breakthroughs in the field of quantum image encryption. Feng {\it et al.}\cite{Feng2022} proposed an image encryption scheme based on the chaotic random behavior characteristics of Boson sampling (BS) probability distribution, which has achieved certain results in many cryptographic applications. Shi {\it et al.}\cite{Shi2022} proposed a novel quantum image encryption scheme based on quantum cellular neural network with quantum operations and hyper-chaotic systems, aiming to optimize security, computation complexity, and decrypted image definition. In 2021, Simon's algorithm was first used to break the 3-round Feistel construction \cite{Kuwakado2010} and proved that the Even-Mansour construction \cite{Kuwakado2012} is insecure with superposition queries. Inspired by them, Kaplan {\it et al.} \cite{Kaplan2016} showed several classical attacks based on finding collisions, which can be greatly accelerated using Simon's algorithm. Shi {\it et al.} \cite{Shi2018} also adopted a similar method to implement a collision attack on the authenticated encrypted AEZ in the CAESAR competition. Recently, according to the parallel and serial structure characteristics of AES-OTR algorithm in processing the associated data, Chang {\it et al.}\cite{Chang2022} constructed periodic function of associated data multiple times based on Simon's algorithm to forge associated data. But as far as we know, there is no quantum forgery attack method for OTR structure ciphertext, and existing quantum attack methods cannot solve this problem. In order to improve the success probability of classical forgery attack on OTR, a quantum forgery attack against OTR structure based on Simon's algorithm is proposed. We conducted quantum forgery attacks directly from the perspective of ciphertext, while Chang {\it et al.}\cite{Chang2022} did so from the perspective of correlated data. Therefore, although the final efficiency is the same, the cost of obtaining information is lower.

In this paper, first, Simon's algorithm is used to obtain the period of the tag generation function in OTR. When the period is obtained, we can forge new messages with known tags. In order to lower the threshold of forgery attack, we propose to conduct forgery attack from the perspective of ciphertext, which makes the attacker only need to intercept the information $(C, T)$ during the communication process between the sender and the receiver, and does not need to know the hard-to-obtain plaintext messages. In addition, for the Prøst-OTR-Even-Mansour structure, we also propose an attack method that only requires some plaintexts to perform a universal forgery attack, which produce the correct ciphertext and tag for any specified message whose ciphertext and tag are not given. Although some conditions are relaxed but it can perform very powerful universal forgery attack. It obtains the secret parameter $L$ using Simon's algorithm and then uses $L$ to obtain keys $k_1$ and $k_2$. The attacker can calculate the tag value for any message, so this attack is the most thorough. The number of queries and the success probability of our quantum forgery attack are mainly reflected in the number of executions of Simon's algorithm and the success probability of finding periods.  That is to say, our quantum forgery attack is not only more realistic, but also has a high probability of success. In addition, the query complexity of our attack is O $(n)$.

The rest of this paper is organized as follows. Sect.2 briefly introduces the OTR authentication encryption algorithm and the Simon’s algorithm. Our quantum forgery attack on the OTR structure based on Simon's algorithm is introduced in Sect.3. Sect.4 specifically presents a universal forgery attack against the Prøst-OTR-Even-Mansour structure. In Sect.5 we give the performance analysis of two attacks. Then  the  conclusions is given in Sect.6.

\section{Preliminaries}
\subsection{OTR structure}
\begin{figure}
\centerline{\includegraphics[width=10cm]{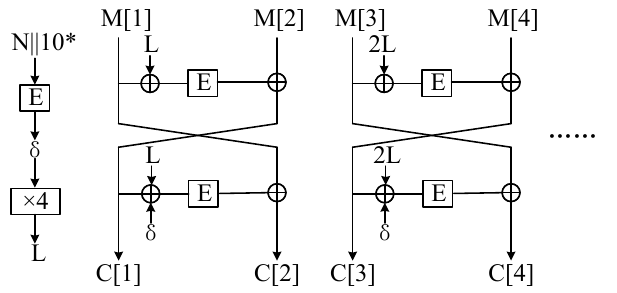}}
\vspace*{8pt}
\caption{The OTR encryption process except for the last set of plaintext\protect\label{Fig1}}
\end{figure}

\begin{figure}
\centerline{\includegraphics[width=14cm]{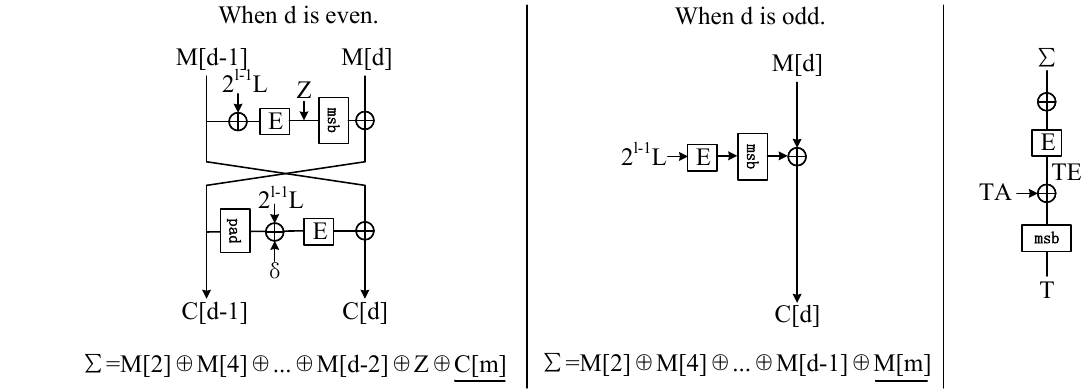}}
\vspace*{8pt}
\caption{The encryption process of the last set of plaintext and the label generation process by OTR\protect\label{Fig2}}
\end{figure}

The OTR structure \cite{Parallelizable2014} accepts the following inputs, key $K \in {\{ 0,1\} ^{\lvert K \lvert} } $, the random number $N\in \{0,1\}^j$ $(1 \le j \le N-1)$, associated data $A\in \{0,1\}^*$ (a binary string of any finite length), plaintext messages $M\in \{0,1\}^*$. And the outputs are ciphertext messages $C\in \{0,1\}^*$ and tag $T\in \{0,1\}^\tau $. The OTR structure divides the plaintext message $M$ into multiple blocks, each containing two plaintext blocks. Then, each block is encrypted using two different masks. These two masks are doubled to get the other two masks for the next block, and so on. The encryption process of OTR removing the last group is shown in Fig. \ref{Fig1}, and the special encryption process of the last group and the part of generating the label are shown in Fig. \ref{Fig2}. The OTR algorithm is described in detail by  Minematsu \cite{Parallelizable2014}.

The authentication token $Tag\_OTR$ in the OTR is obtained by  $TE$ XOR $TA$ ($Tag\_OTR = TE \oplus TA$). If the associated data $A$ is equal to the empty string, then the final tag $Tag\_OTR$ will be equal to $TE$.

\subsection{Simon's algorithm}
Simon's problem: Given a boolean function $f:{\{ 0,1\} ^n} \to  {\{ 0,1\} ^n}$, there exists $s \in {\{ 0,1\} ^n}$, such that  $f\left( x \right){\rm{ }} = {\rm{ }}f\left( y \right){\rm{ }}$ for all ${\rm{ }}(x,{\rm{ }}y) \in {\{ 0,1\} ^{2n}}$, where $x \oplus y \in \{ {0^n},{\rm{ }}s\} $, and the goal is to find $s$.

This problem can be solved by looking for collisions. So the best time to solve it is $\Theta (2^{n/2})$. On the other hand, Simon's algorithm \cite{Simon1997} solves this problem with quantum complexity $O(n)$, and it repeats the following five steps.

\begin{romanlist}[(123)]
 \item Initialized with $2n$ qubits $\left| 0 \right\rangle\left| 0 \right\rangle$, one of the registers applies the Hadamard transformation ${\rm{H}} ^{\otimes {\rm{n }}}$ to obtain a quantum superposition.
 \begin{equation}
     \frac{1}{{\sqrt {{2^n}} }}\sum\limits_{x \in {{\{ 0,1\} }^n}} {\left| x \right\rangle } \left| 0 \right\rangle 
 \end{equation}
 \item A quantum query on a function $f$ maps it to the state,
 \begin{equation}
     \frac{1}{{\sqrt {{2^n}} }}\sum\limits_{x \in {{\{ 0,1\} }^n}} {\left| x \right\rangle } \left| f(x) \right\rangle 
 \end{equation}
\item Measure the second register to get a value $ f(z)$ based on the calculation and fold the first register to the state,
 \begin{equation}
     \frac{1}{{\sqrt {{2^n}} }}(\left| z \right\rangle  + \left| {z \oplus s} \right\rangle ) 
 \end{equation}
 \item Applying the Hadamard transformation $H^{\otimes n}$ again to the first register yields,
 \begin{equation}
    \frac{1}{{\sqrt 2 }}\frac{1}{{\sqrt {{2^n}} }}\sum\limits_{y \in {{\{ 0,1\} }^n}} {{{( - 1)}^{y \cdot z}}} (1 + {( - 1)^{y \cdot z}})\left| y \right\rangle 
 \end{equation}
 \item The vectors $y$ such that $y\cdot s =1$ has an amplitude of 0. Therefore, measuring the state in the computational base yields a random vector $y$ such that $y\cdot s = 0$.
\end{romanlist}

By repeating this subroutine $O(n)$ times, one obtains $n-1$ independent vectors with high probability orthogonal to $s$, which can be recovered with basic linear algebra.

\section{Quantum Forgery Attack on OTR Using Simon's Algorithm}
Our attack excludes the interference of the associated data, that is, setting the associated data equal to the empty string. Now, according to the label calculation formula $T=TE$, we found that if $d\ge 4$, then Simon's algorithm can be used to find the period value of $TE$. Because of some characteristics of OTR tail processing, the forgery results are different when $d=4$ and $d>4$. So, we discuss it in two cases below. First let's look at the general case, if $d>4$, then according to the OTR encryption algorithm, we can know that, 
\begin{equation}
    TE = E\left( {3{L^*} \oplus \delta  \oplus M[2] \oplus M[4] \oplus ... \oplus M[d]} \right)
\end{equation}
where $L^{*}=L \oplus \delta$. When $d>4$, suppose $\left| {M[{\rm{d}}]} \right| = n$, then
\begin{equation}
     M[2] = E\left( {4\delta  \oplus E\left( {5\delta  \oplus C[1]} \right) \oplus C[2]} \right) \oplus C[1]
 \end{equation}
  \begin{equation}
     M[4] = E\left( {8\delta  \oplus E\left( {9\delta  \oplus C[3]} \right) \oplus C[4]} \right) \oplus C[3]
 \end{equation}

By substituting  Eq.6 and  Eq.7 into Eq.5, we can get the relevant formulas of the ciphertext and $TE$, as follows,
\begin{equation}
\begin{split}
 TE = E(3{L^*} \oplus \delta  \oplus E( {4\delta  \oplus E({5\delta  \oplus C[1]}) \oplus C[2]} )\oplus C[1] \oplus\\
 E( {8\delta  \oplus E( {9\delta  \oplus C[3]} ) \oplus C[4]} ) \oplus C[3] \oplus ... \oplus M[d])
\end{split}
\end{equation}

By Eq.8, we find that whether $d$ is odd or even and whether the length of the last block of plaintext is $n$, our falsification is only related to $C[1], C[2], C[3], C[4]$, so we assume $d=5$, then
\begin{equation}
\begin{split}
TE = E(3{L^*} \oplus \delta  \oplus E( {4\delta  \oplus E( {5\delta  \oplus C[1]} ) \oplus C[2]} ) \oplus C[1] \oplus \\
E( {8\delta  \oplus E( {9\delta  \oplus C[3]} ) \oplus C[4]} ) \oplus C[3] \oplus E(16\delta ) \oplus C[5])
\end{split}
\end{equation}

\begin{figure}
\centerline{\includegraphics[width=8cm]{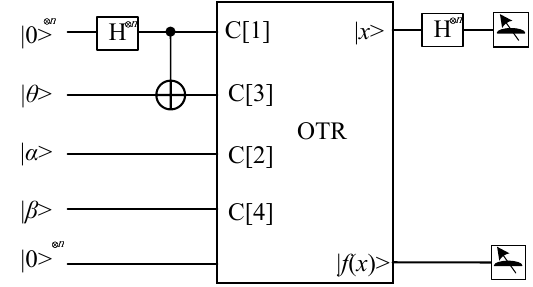}}
\vspace*{8pt}
\caption{Quantum forgery attack circuit on OTR\protect\label{Fig3}}
\end{figure}

We define the following function:
\begin{equation}
\begin{aligned}
\begin{array}[array]{rll}
f_{a}:{\{ {0,1} \}^n} &\to {\{ {0,1} \}^n}& \\
x&  \to Tag\_OTR( {x\lvert \lvert x \oplus \theta ,\alpha \lvert \lvert \beta } )  &\\ &=  E( 3{L^*} \oplus \delta  \oplus E( {4\delta  \oplus E( {5\delta  \oplus x} ) \oplus \alpha } )     \oplus E( {8\delta  \oplus E(
{9\delta  \oplus x \oplus \theta } ) \oplus \beta } ) \oplus \theta  \oplus E( {16\delta } ) \oplus C[5] )
\end{array}
\end{aligned}
\end{equation}
where $\theta  = C[1] \oplus C[3]$, and $\alpha$,$\beta $ ($C[2]$,$C[4]$) are constants. We only need one call to the cryptographic oracle to complete the construction of the function $f_{a}$. A quantum circuit is built for $f$, as shown in Fig. \ref{Fig3}. We find that $f_{a}$ satisfies the requirements of Simon's algorithm. It is obvious to see that
$f_{a}\left( x \right) = f_{a}\left( {x \oplus s} \right)$ with $s = 13\delta  \oplus \theta$.

\begin{equation}
\begin{split}
f_{a}(x) = E( 3{L^*} \oplus \delta  \oplus E( {4\delta  \oplus E( {5\delta  \oplus x} ) \oplus \alpha } ) \oplus  E( 8\delta\\  \oplus 
E( 9\delta  \oplus x \oplus \theta ) \oplus \beta  ) \oplus \theta  \oplus E( {16\delta } ) \oplus C[5])
\end{split}
\end{equation}

\begin{equation}
\begin{aligned}
\begin{array}[array]{rl}
f_{a}({x\oplus s} ) &= E(3{L^*} \oplus \delta  \oplus E( 4\delta  \oplus E( 5\delta  \oplus x \oplus s ) \oplus \alpha  )\oplus E( 8\delta  \oplus E( 9\delta  \oplus x \oplus s \oplus \theta  ) \\ 
& \oplus \beta  ) \oplus \theta  \oplus E(16\delta  ) \oplus C[5] )\\
& = E( 3{L^*} \oplus \delta  \oplus E( 4\delta  \oplus E( 5\delta  \oplus x \oplus 13\delta  \oplus \theta  ) \oplus \alpha  ) \oplus E( 8\delta  \oplus E(9\delta  \oplus x \oplus 13\delta  \oplus \theta  \oplus \theta  )\\
&\oplus \beta  ) \oplus \theta  \oplus E( 16\delta ) \oplus C[5] )\\
&= E( 3{L^*} \oplus \delta  \oplus E( 4\delta  \oplus E(9\delta  \oplus x \oplus \theta ) \oplus \alpha  )\oplus E( 8\delta  \oplus E( 5\delta  \oplus x ) \oplus \beta  ) \oplus \theta  \oplus E(16\delta  ) \oplus C[5] )\\
&= f_{a}(x)
\end{array}
\end{aligned}
\end{equation}

We set $C[1]$ and $C[3]$ as two constants $\chi $ and $\varpi $,  then $E\left( {5\delta  \oplus C[1]} \right)$ and $E\left( {9\delta  \oplus C[3]} \right)$ also as two constants (both are set $\mu$) and do not affect the period value. The function can also be defined through the tag generation function,
\begin{equation}
\begin{aligned}
\begin{array}[array]{rl}
f_{b}:{\left\{ {0,1} \right\}^n} & \to {\left\{ {0,1} \right\}^n}\\
x & \to Tag\_OTR\left( {x||x \oplus \tau , \chi ||\varpi ||\mu } \right)\\
& = E\left( {3{L^*} \oplus \delta  \oplus E\left( {4\delta  \oplus \mu  \oplus x} \right) \oplus \chi  \oplus E\left( {8\delta  \oplus \mu  \oplus x \oplus \tau } \right) \oplus \varpi  \oplus E\left( {16\delta } \right) \oplus C[5]} \right)
\end{array}
\end{aligned}
\end{equation}
where $\tau = C[2] \oplus C[4]$. It is obvious to see that
${f_{b}}\left( x \right) = f_{b}\left( {x \oplus s} \right)$ with $s = 12\delta  \oplus \tau $.

\begin{equation}
\begin{split}
{f_{b}}\left( x \right) = E\left( {3{L^*} \oplus \delta  \oplus E\left( {4\delta  \oplus \mu  \oplus {x}} \right) \oplus \chi 
\oplus E\left( {8\delta  \oplus \mu  \oplus x \oplus \tau } \right) \oplus \varpi  \oplus E\left( {16\delta } \right) \oplus C[5]} \right)
\end{split}
\end{equation}

\begin{equation}
\begin{aligned}
\begin{array}[array]{rl}
{f_{b}}( x \oplus {\rm{s}} ) & = E( 3{L^*} \oplus \delta  \oplus E(4\delta  \oplus \mu  \oplus {x} \oplus {s} ) \oplus \chi 
\oplus E( 8\delta  \oplus \mu  \oplus x \oplus {s} \oplus \tau  )\\
& \oplus \varpi  \oplus E( {16\delta } ) \oplus C[5] )\\
& = E( 3{L^*} \oplus \delta  \oplus E(4\delta  \oplus \mu  \oplus {x} \oplus 12\delta  \oplus \tau  ) \oplus \chi \oplus E(8\delta  \oplus \mu  \oplus x \oplus 12\delta  \oplus \tau  \oplus \tau )\\
& \oplus \varpi  \oplus E(16\delta) \oplus C[5] )\\
& = E(3{L^*} \oplus \delta  \oplus E(8\delta  \oplus \mu  \oplus {x} \oplus \tau ) \oplus \chi \oplus E(4\delta  \oplus \mu  \oplus x ) \oplus \varpi  \oplus E(16\delta ) \oplus C[5] )\\
& = {f_{b}}( x )
\end{array}
\end{aligned}
\end{equation}
So, the tag $T$ of $C=C[1]\lvert\lvert C[2] \lvert\lvert C[3] \lvert\lvert C[4] \lvert\lvert ... \lvert\lvert C[d] $ is also valid for $C'= 13\delta  \oplus C[3] \lvert\lvert 12\delta  \oplus C[4] \lvert\lvert 13\delta  \oplus C[1] \lvert\lvert 12\delta  \oplus C[2] \lvert\lvert ...\lvert\lvert C[d] $, where $T$ was intercepted during communication process.

When $d=4$, because of the special handling of parity at the end of the OTR, the function will be slightly different from the function design of $d>4$. Now,

\begin{equation}
\begin{split}
TE = E\left( {26\delta  \oplus E\left( {4\delta  \oplus E\left( {5\delta  \oplus C[1]} \right) \oplus C[2]} \right) \oplus C[1] \oplus E\left( {8\delta  \oplus E\left( {9\delta  \oplus C[4]} \right) \oplus C[3]} \right) \oplus C[4]} \right)
\end{split}
\end{equation}

It is not difficult to find that when $d=4$, the positions of $C[4]$ and $C[3]$ are swapped in the function. It can be deduced that when $d>4$, the tag of $C = C[1] \lvert\lvert C[2] \lvert\lvert C[3]\lvert\lvert C[4]\lvert\lvert ...\lvert\lvert C[d]$ is also works for $C'=13\delta  \oplus C[4] \lvert\lvert  12\delta  \oplus C[3] \lvert\lvert 12\delta  \oplus C[2] \lvert\lvert 13\delta  \oplus C[1] \lvert\lvert ......\lvert\lvert C[d]$. It is worth noting that due to space limitations, the forgery of $d \leq 4$ is not introduced in this example. And in the encryption process, due to the limitation of the calculation structure of the tag, the OTR structure is not suitable using Simon’s algorithm for quantum forgery attack when $d \leq 4$.

\section{Quantum Forgery Attack on Prøst-OTR-Even-Mansour}
\begin{figure}[ph]
\centerline{\includegraphics[width=10cm]{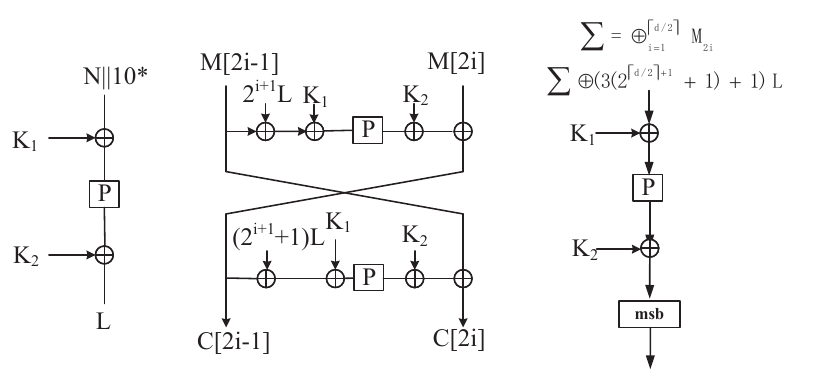}}
\vspace*{8pt}
\caption{Core part of the Prøst-OTR-Even Mansour structure\protect\label{Fig4}}
\end{figure}

The Prøst-OTR authenticated encryption algorithm is a CAESAR candidate submitted by Kavun et al. \cite{Prøstv1,Prøstv1.1}, which is inspired by OTR and combines new efficient permutations. Its variant uses the Prøst permutation in the Even Mansour structure for cryptographic operations, and we call this variant the Prøst-OTR-Even-Mansour structure. As shown in the Fig. \ref{Fig4}, because the associated data has no effect on our attack, we assume that the associated data $A$ is an empty string, and we won't introduce it here.

Prøst-OTR-Even-Mansour still retains the special calculation method of the OTR structure for the tag, that is, the calculation method of the tag is affected by the parity of $d$ and the length of the last plaintext block. For convenience, we first assume that $\left| {M[d]} \right| = n$, then
\begin{equation}
\begin{split}
TE = {k_2} \oplus P\left( {\Sigma  \oplus \left( {3\left( {{2^{\left\lceil {d/2} \right\rceil  + 1}+1}} \right) + 1} \right)L \oplus {k_1}} \right).
\end{split}
\end{equation}
where $\Sigma  = M[2] \oplus M[4] \oplus ... \oplus M[d]$. Note that almost all forgery attacks allow a given message $M$ to be known to the forger. Through Eq.17, we find that if the parameter $L$ can be known, Simon's algorithm can be used to carry out a very powerful universal forgery attack on this structure. The following method is proposed by us to obtain the parameter $L$.

First we intercept the plaintext $d=2$ and $4$. If $d=2$, we can get Eq.18, and if $d=4$, then we can get Eq.19.
\begin{equation}\label{eq18}
\begin{split}
TE = {k_2} \oplus P\left( {{\rm{M}}[2] \oplus 16L \oplus {k_1}} \right)
\end{split}
\end{equation}
\begin{equation}\label{eq19}
\begin{split}
TE = {k_2} \oplus P\left( {{\rm{M}}[2] \oplus {\rm{M}}[4] \oplus 26L \oplus {k_1}} \right)
\end{split}
\end{equation}

Next, we set $M[2]$ as the input $x$ and $M[4]$ as the constant $c$, then we can construct the function $f_{c}$:
\begin{equation}\label{eq20}
\begin{aligned}
\begin{array}[array]{rl}
f_{c}:{\left\{ {0,1} \right\}^n} & \to {\left\{ {0,1} \right\}^n}\\
x & \to Tag\_OTR\left( x \right) \oplus Tag\_OTR\left( {x||{\rm{c}}} \right) \\
& = P\left( {x \oplus 16L \oplus {k_1}} \right) \oplus P\left( {x \oplus c \oplus 26L \oplus {k_1}} \right)
\end{array}
\end{aligned}
\end{equation}
Now, we can apply Simon algorithm on this function $f_{c}$.  It is obvious to see that $f_{c}\left( x \right) = f_{c}\left( {x \oplus s} \right)$ with $s = c \oplus 10L$.
\begin{equation}
\begin{split}
f_{c}\left( x \right) = P\left( {x \oplus 16L \oplus {k_1}} \right) \oplus P\left( {x \oplus c \oplus 26L \oplus {k_1}} \right)
\end{split}
\end{equation}

\begin{equation}
\begin{aligned}
\begin{array}[array]{rl}
f_{c}\left( {x \oplus s} \right) & = P\left( {x \oplus s \oplus 16L \oplus {k_1}} \right) \oplus P\left( {x \oplus s \oplus c \oplus 26L \oplus {k_1}} \right)\\
 &= P\left( {x \oplus c \oplus 10L \oplus 16L \oplus {k_1}} \right) \oplus P\left( {x \oplus c \oplus 10L \oplus c \oplus 26L \oplus {k_1}} \right)\\
& = P\left( {x \oplus c \oplus 26L \oplus {k_1}} \right) \oplus P\left( {x \oplus 16L \oplus {k_1}} \right)\\
& = f_{c}\left( x \right)
\end{array}
\end{aligned}
\end{equation}

We can find the parameters $L$ through $10L = s \oplus c$. Now, $\Sigma  \oplus \left( {3\left( {{2^{\left\lceil {d/2} \right\rceil  - 1}}} \right) + 1} \right)L$ can be used as input $x$. $f_{d}$ can be constructed as,

\begin{equation}
\begin{aligned}
\begin{array}[array]{rl}
f_{d}:{\left\{ {0,1} \right\}^n} & \to {\left\{ {0,1} \right\}^n}\\
x & \to Tag\_OTR\left( x \right) \oplus P\left( x \right) = {k_2} \oplus P\left( {x \oplus {k_1}} \right) \oplus P\left( x \right)
\end{array}
\end{aligned}
\end{equation}
where $P$ is the Prøst permutation operation, which is public. When $s = {k_1}$, $f_{d}\left( x \right) = f_{d}\left( {x \oplus s} \right)$ can be obtained by Simon's algorithm. Knowing the key ${k_1}$, we can calculate the key ${k_2}$ in the following way.

\begin{equation}
\begin{aligned}
\begin{array}[array]{rl}
{k_2} & = P\left( {x \oplus {k_1}} \right) \oplus Tag\_OTR\left( x \right) = P\left( {x \oplus {k_1}} \right) \oplus {k_2} \oplus P\left( {x \oplus {k_1}} \right)
\end{array}
\end{aligned}
\end{equation}

Now that we have the key ${k_1}$ and ${k_2}$, and we can perform a universal forgery attack, in which the attacker can calculate the ciphertext $(C)$ and tag value $(T)$ for any message. So that the attacker can send $(C,T)$ of a forge message to the receiver, and the receiver cannot identify the sender. Also when $d$ is an even number and $\left| {M[d]} \right| = n$, we find that the parity affects the calculation method of $\Sigma$ (see Fig. \ref{Fig2}), but this does not affect our input and label calculation method. It is also feasible to use Simon's algorithm to attack. For the length of the last block of plaintext, OTR adopts pad zero-filling method, which also has no effect on the feasibility of our attack. To sum up, regardless of the value of $d$ and the length of the last block of plaintext, we can perform a universal forgery attack.

\section{Performance analysis}

In this section, we analyze the performance of the two attacks proposed in Sect.3 (Attack 1) and Sect.4 (Attack 2) for two different structures. Since the attacks only require several blocks of ciphertext or plaintext ($d>4$), its contribution on performance can be ignored. Therefore, our analysis mainly includes success probability, query complexity and qubit number.

(1) Attacks analysis on OTR structure

Zheng’s attack \cite{Zheng2017} proposed a forgery attack method in the case of only knowing one pair of plaintext-ciphertext (Zheng's attack 1) and multiple pairs of plaintext-ciphertext (Zheng's attack 2) for the OTR structure. They require $O(n^2)$ queries and $O(n^r)$ queries, respectively. The success probability of two situations is $r^{2}·2^{-n}$ and $s^{2}(s+1)2^{r-n}$ respectively, where $r=(d-1)/2$, $n$ is the size of the block and $s+1$ is the number of known plaintext and ciphertext pairs \cite{Zheng2017}. It can be seen that $2^n$ ($n$ is usually taken as 128) is a very large number, so their success probability has room for improvement.

\begin{figure}[ph]
\centerline{\includegraphics[width=10cm]{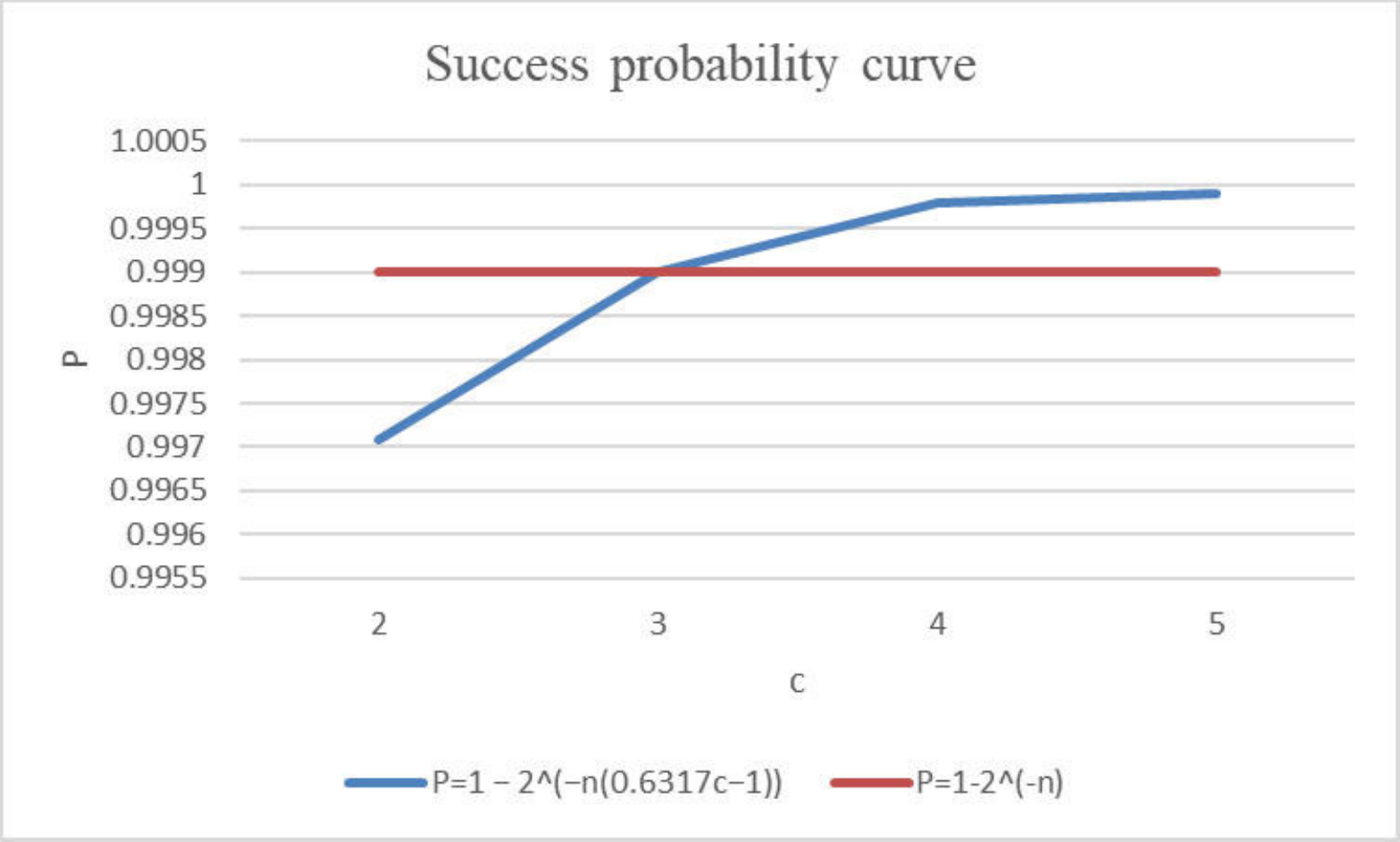}}
\vspace*{8pt}
\caption{Success probability curve\protect\label{Fig5}}
\end{figure} 

Attack 1 only needs to know some ciphertexts that are about to be forged. As an application of Simon's algorithm, Attack 1's success probability may be slightly lower than that of the strict Simon's algorithm, since the OTR structure does not necessarily meet the strict Simon's problem. In 2019, Shi {\it et al}.  proved that even if the condition of Simon's algorithm is not strictly satisfied, it will still return the correct result with $cn$ queries, with probability $P \ge 1-2^n  \times  (0.6454)^{cn}$ \cite{Shi2019}. If $c \geq 4$, then the success probability of Attack 1 is
\begin{equation}
\begin{aligned}
\begin{array}[array]{rl}
P& =1-2^{n}\times(0.6454)^{cn} \\
& =1-2^{-n(-c\log_2{(0.6454)}-1)} \\
& \approx 1-2^{-n(0.6317c-1)}\\
& \ge 1-2^{-n},
\end{array}
\end{aligned}
\end{equation}
which easily approaches 1 and does not depend on the query complexity \cite{Shi2019}. The success probability curve is shown in the Fig. \ref{Fig5}. Therefore, its query complexity is $cn=O(n)$, and the qubit number is $O(n)$. 
The attack comparison of OTR is shown in Table \ref{ta1}.

\begin{table}[h]
\tbl{Attack comparison of OTR.}
{\begin{tabular}{@{}ccccc@{}} \toprule
Attack & Plaintext(P)/Ciphertext(C) & Query & Success probability $P$ \\
\colrule
Zheng's attack 1 \cite{Zheng2017} & P+C & $O(n^2)$ & $r^{2}·2^{-n}$  \\
Zheng's attack 2 \cite{Zheng2017} & P+C & $O(n^r)$ &  $s^{2}(s+1)2^{r-n}$ 
\\
Attack 1 & C & $O(n)$ & $1-2^n  \times  (0.6454)^{cn}$  \\
\botrule
\end{tabular}\label{ta1} }
\end{table}

As shown in Table \ref{ta1}, Zheng's attack 1 and 2 need plaintext-ciphertext pairs, while Attack 1 only need some ciphertexts. It is obvious that Attack 1's scenario is more realistic. We also have the lowest query complexity $O(n)$. The success probability of Attack 1 is higher than Zheng's attack 1 and 2.

(2) Attacks analysis on Prøst-OTR-Even-Mansour structure

Christoph's attack \cite{Christoph2015} needs to give the ciphertext and label of any two messages under two related keys, and then they can forge and modify the ciphertext and label of the message. Their attack is selective forgery attack. It has $2^{-n/2}$ success probability \cite{Saarinen2014}. The query complexity is not discussed because it is a related-key attack, but the query complexity of Attack 2 is not high either, it is $O(n)$. Attack 2 only need some plaintexts and the attacker can forge all the messages, which is a universal forgery attack. Like Attack 1, Attack 2 requires $O(n)$ queries and has a success probability of $1-2^n  \times  (0.6454)^{cn}$.
That is to say, our quantum forgery attack not only has a loose scenario but also has a high success probability. The attack comparison of Prøst-OTR-Even-Mansour is shown in Table \ref{ta2}.

\begin{table}[h]
\tbl{Attack comparison of Prøst-OTR-Even-Mansour.}
{\begin{tabular}{@{}ccccc@{}} \toprule
Attack & Plaintext(P)/Ciphertext(C) & Success probability $P$ & Attack type \\
\colrule
Christoph's attack \cite{Christoph2015} & P+C &  $2^{-n/2}$ & Selective forgery attack \\
Attack 2 & P  &  $1-2^n  \times  (0.6454)^{cn}$ & Universal forgery attack \\
\botrule
\end{tabular}\label{ta2} }
\end{table}

\section{Conclusion}
In this paper, based on Simon's algorithm, two forgery attacks (quantum forgery attack on OTR structure and quantum forgery attack on Prøst-OTR-Even-Mansour structure) are proposed. The former attack is based on ciphertext and the attacker does not know the content of the message and can interfere with message transmission. Different from the former attack, the latter attack is a universal forgery attack, which is based on several plaintext blocks. The attacker can obtain the content of the messages, and forge the messages purposefully. In view of this, we believe that quantum forgery attacks have a great advantage in other areas of authenticated encryption modes. However, if the attacker only queries classically, our quantum forgery attack will lose the attack premise. Therefore, we consider using quantum algorithms offline to
improve the efficiency of breaking authenticated encryption modes, which will be one of our future research directions.

\section*{Acknowledgments}
This work is supported by the National Natural Science Foundation of China (62071240), the Innovation Program for Quantum Science and Technology (2021ZD0302902), and the Priority Academic Program Development of Jiangsu Higher Education Institutions (PAPD).


\end{document}